\documentclass[twocolumn,showpacs,preprintnumbers,amsmath,amssymb]{revtex4}

\usepackage{graphicx}
\usepackage{dcolumn}
\usepackage{bm}
\usepackage{colordvi}

\begin{document}


\title{Static and Dynamic Correlations in Many-Particle Lyapunov Vectors}
\author{G\"unter Radons}%
 \email{radons@physik.tu-chemnitz.de}
\author{Hongliu Yang}
 \email{hongliu.yang@physik.tu-chemnitz.de}
\affiliation{%
Institut f\"ur Physik, Theoretische Physik I,
Technische Universit\"at Chemnitz, D-09107 Chemnitz, Germany 
}%

\begin{abstract}
We introduce static and dynamic correlation functions for the spatial
densities of Lyapunov vector fluctuations. They enable us to show, for the
first time, the existence of hydrodynamic Lyapunov modes in chaotic
many-particle systems with soft core interactions, which indicates
universality of this phenomenon. Our investigations for Lennard-Jones fluids
yield, in addition to the Lyapunov exponent - wave vector dispersion, the
collective dynamic excitations of a given Lyapunov vector. In the limit of
purely translational modes the static and dynamic structure factor are
recovered.
\end{abstract}

\pacs{05.45.-a, 05.20.-y, 02.70.Ns}
\maketitle




The dynamical instability of systems, which is quantified by the spectrum of
Lyapunov exponents, has long been recognized as a most important
characteristic for chaotic systems \cite{EckmannRuelle}. Its understanding
for extended dynamical systems provides essentials for the foundations of
equilibrium and non-equilibrium statistical mechanics \cite{Dorfman}. This
explains the on-going scientific activities aiming at the connection between
Lyapunov spectra of many-particle systems and macroscopic properties such as
transport coefficients \cite{vBeijeren} or entropy production rates \cite%
{DellagoPoschHoover96,ParingRulesEntropyProduction,PoschForster}. To each of
the Lyapunov exponents in the spectrum one associates a Lyapunov vector in
state or phase space, which in contrast to the Lyapunov exponent is a
time-dependent quantity. Apart from singular early investigations \cite%
{Pomeau} these mutually orthogonal vectors were for a long time used mainly
as a means for calculating Lyapunov spectra. The scientific interest in
these quantities raised when it was realized that especially in extended
systems they contain non-trivial information about the perturbation
dynamics. For instance, for certain coupled map lattices it was found that
the Lyapunov vectors are often localized in state space \cite{LVlocalizedCML}%
, a fact that was later also observed in many-particle systems \cite%
{LVlocalizedMPS,ForsterHirschlPoschHoover,PoschForster}. Very recently,
however, a very surprising phenomenon was observed. In contrast to the
localized behavior of vectors belonging to the largest Lyapunov exponents,
it was found that the Lyapunov vectors corresponding to near-zero exponents
exhibit wave-like structures \cite%
{LVhydrodynamics,ForsterHirschlPoschHoover,PoschForster}. Due to the
approximately linear relation between Lyapunov exponent and wave number of
the associated vector the latter were termed hydrodynamic Lyapunov modes.
These findings immediately triggered theoretical efforts to gain deeper
insights on the basis of simplified models such as products of random
matrices \cite{LModeTheory}. However, a really satisfactory state of
understanding has not yet been achieved. For instance, molecular dynamics
simulations indicated that hydrodynamic Lyapunov modes seem to exist only in
hard-sphere or hard-disk fluids, but not in many-particle systems with soft
interaction potentials \cite%
{LVhydrodynamics,ForsterHirschlPoschHoover,PoschForster}. Although there are
some arguments for this distinctive behavior based on fluctuations of finite
time Lyapunov exponents, the existence of such a sharp difference between
hard and soft potentials remains surprising. The purpose of this Letter is
to show that hydrodynamic Lyapunov modes actually do exist also in
soft-potential many-particle systems such as Lennard-Jones (LJ) fluids. This
is achieved by considering static and dynamic correlation functions for the
density of Lyapunov vector fluctuations, which we define appropriately in
the spirit of generalized hydrodynamics \cite{GeneralizedHydrodynamics}.

Before we present our results for Lennard-Jones fluids let us briefly recall
the definition of Lyapunov vectors and their dynamics. For an autonomous
dynamical system in $\mathbb{R}^{D}$ given by $\dot{\Gamma}=G(\Gamma )$ the
tangent space dynamics describing small perturbation around a reference
trajectory $\Gamma (t)$ is given by $\delta \dot{\Gamma}=M(\Gamma (t))\cdot
\delta \Gamma $ with the Jacobian $M=\frac{dG}{d\Gamma }$. For a $D-$%
dimensional dynamical system there exist $D$\ Lyapunov vectors $e^{(\alpha
)}(t)\in \mathbb{R}^{D}$, $\alpha =1,\ldots ,D$ which obey the equations of
motion \cite{Goldhirsch,HooverPoschBestiale}%
\begin{equation}
\dot{e}^{(\alpha )}=M(\Gamma (t))\cdot e^{(\alpha )}-\sum_{\beta =1}^{\alpha
}m_{\alpha \beta }(t)~e^{(\beta )}  \label{LVdynamics}
\end{equation}%
with $m_{\alpha \beta }(t)=\left[ e^{(\alpha )}\cdot M(\Gamma (t))\cdot
e^{(\beta )}+e^{(\beta )}\cdot M(\Gamma (t))\cdot e^{(\alpha )}\right] $ $%
\times (1-\delta _{\alpha \beta }/2)$. Eq.(\ref{LVdynamics}) describes the
dynamics of a frame consisting of $D$ perturbation vectors $\delta \Gamma $
which are kept orthonormal by the nonlinear terms containing the Lagrange
multipliers $m_{\alpha \beta }(t)$. The Lyapunov exponents $\lambda
^{(\alpha )}$ are given by the time averages $\lambda ^{(\alpha
)}=\left\langle m_{\alpha \alpha }(t)\right\rangle _{t}$, with $\left\langle
\ldots \right\rangle _{t}=\lim_{t\rightarrow \infty }\frac{1}{t}%
\int_{0}^{t}\ldots dt^{\prime }$. In practice one usually does not integrate
Eq.(\ref{LVdynamics}) because of numerical instabilities, but rather obtains
the Lyapunov vectors $e^{(\alpha )}(t)$ (and Lyapunov exponents) by the
so-called standard method consisting of repeated Gram-Schmidt
orthogonalization of an evolving set of $D$ perturbation vectors $\delta
\Gamma (t)$ \cite{EckmannRuelle}. One should note that with this procedure
the Lyapunov vectors are automatically ordered such that $e^{(1)}$ belongs
to the largest Lyapunov exponent $\lambda ^{(1)}$, $e^{(2)}$ to the second
largest $\lambda ^{(2)}$, and so on. Eq.(\ref{LVdynamics}) corresponds to
the continuum limit of this procedure and is given here explicitly to
emphasize that $e^{(\alpha )}(t)$ is a well defined dynamical variable in
continuous time. 
\begin{figure}[tbp]
\includegraphics*[scale=0.24]{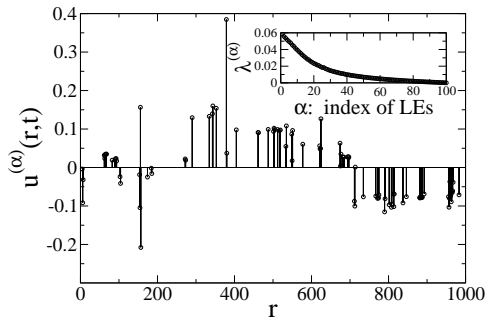}
\caption{Snapshot of the LV density $u^{(\protect\alpha )}(r,t)$ for a 1-d
Lennard-Jones fluid ($N=100$) for which the corresponding Lyapunov exponent $%
\protect\lambda ^{(96)}$ is close to zero. The displacements $u_{i}^{(%
\protect\alpha )}$ drawn vertically at the current particle positions $R_{i}$
show indications of a wave-like structure. The insert shows  for this system
the positive branch of the Lyapunov spectrum.}
\label{fig:1}
\end{figure}

We treat the Hamiltonian dynamics of $N$ interacting particles of a $d-$%
dimensional fluid. Therefore $\Gamma =(R,P)$ is a phase space vector in $%
D=2dN$ dimensions evolving as $\Gamma (t)=\exp (i\mathcal{L}t)~\Gamma (0)$
where $\mathcal{L}$ denotes the Liouville operator. The coordinate part of $%
\Gamma $ is\ written as $R=(R_{1},\ldots ,R_{N})$, with $R_{i}\in \mathbb{R}%
^{d}$ being the coordinate vector of the $i-$th particle. The momentum part $%
P$ is defined analogously. Correspondingly the Lyapunov vectors $e^{(\alpha
)}$ can also be split into a coordinate and a momentum part as $e^{(\alpha
)}=(u_{1}^{(\alpha )},\ldots ,u_{N}^{(\alpha )},v_{1}^{(\alpha )},\ldots
,v_{N}^{(\alpha )})$. In the following we will concentrate on correlations
in the coordinate part of these vectors. For this purpose we introduce the
dynamical variable $u^{(\alpha )}(r,t)=\sum_{i=1}^{N}u_{i}^{(\alpha
)}(t)~\delta (r-R_{i}(t))$, which describes the spatial density of the
coordinate part of Lyapunov vector fluctuations, which we call for short LV
density. This variable is defined in analogy to other microscopic density
fluctuations, e.g. of the particle current, in molecular hydrodynamics \cite%
{GeneralizedHydrodynamics}. A snapshot of this dynamical variable is shown
in Fig.1 for a 1-d LJ fluid. For translational invariant systems as treated
here, it is convenient to consider the correlations of its Fourier transform 
$u^{(\alpha )}(k,t)=\int u^{(\alpha )}(r,t)~\exp (-ikr)~dr$ given by

\begin{equation}
u^{(\alpha )}(k,t)=\sum_{i=1}^{N}u_{i}^{(\alpha )}(t)~\exp (-ikR_{i}(t))
\label{LVFourier}
\end{equation}%
The intermediate two-time correlation function of this variable is defined
as 
\begin{equation}
F^{(\alpha \alpha )}(k,\tau )=N^{-1}\left\langle u^{(\alpha )}(k,t+\tau
)~u^{(\alpha )}(-k,t)\right\rangle _{t}  \label{intermediate}
\end{equation}%
which for $d\neq 1$ is a second rank tensor. From the latter we get for $%
\tau =0$ the static LV density correlation function $S^{(\alpha \alpha
)}(k)=F^{(\alpha \alpha )}(k,\tau =0)$ and by Fourier transformation with
respect to time the dynamical LV density correlation function $S^{(\alpha
\alpha )}(k,\omega )=(2\pi )^{-1}\int F^{(\alpha \alpha )}(k,\tau )\exp
(i\omega \tau )~d\tau $. Note that in contrast to the Lyapunov vectors
itself, their correlation functions, being time averages, define \emph{global%
} quantities similar to the Lyapunov exponents $\lambda ^{(\alpha )}$. Even
more far reaching is the following fact. It has been shown \cite{Ershov}
that generically for long times the Lyapunov vectors $e^{(\alpha )}(t)$
depend on time only via its \emph{current} phase space point $\Gamma (t)$,
especially it does not depend on the initial conditions $e^{(\alpha )}(t=0)$
as Eq.(\ref{LVdynamics}) might suggest. This implies that in long time
averages $e^{(\alpha )}(t)$ can be replaced by $e^{(\alpha )}(\Gamma (t))$,
where the vector fields $e^{(\alpha )}(\Gamma )$, $\alpha =1,\ldots ,D$,
provide what in \cite{Ershov} is called a ''stationary Lyapunov basis''.
Accordingly the evolution of the dynamical variables, e.g. Eq.(\ref%
{LVFourier}), can be expressed as $u^{(\alpha )}(k,t)=\exp (i\mathcal{L}%
t)~u^{(\alpha )}(k,0)$ and, conceptually most important, the time averages
in our correlation functions can be evaluated also as ensemble averages $%
F^{(\alpha \alpha )}(k,\tau )=N^{-1}\left\langle u^{(\alpha
)}(k,t)~u^{(\alpha )}(-k,0)\right\rangle $ over the invariant density $\mu
(\Gamma )$, i.e. $\left\langle \ldots \right\rangle =\int \ldots \mu (\Gamma
)~d\Gamma $. Furthermore standard techniques for the analytical evaluation
of these correlation functions, such as the Mori-Zwanzig projection operator
formalism \cite{GeneralizedHydrodynamics}, could be applied in principle.
The problem, however, is the lack of knowledge about the vector fields $%
e^{(\alpha )}(\Gamma )$. One of the few known features is that for systems
with translational invariance there exist Lyapunov vectors with constant
spatial component $u_{i}^{(\alpha _{0})}=c^{(\alpha _{0})}$ belonging to the
zero Lyapunov exponents $\lambda ^{(\alpha _{0})}=0$ related to this
symmetry \cite{LModeTheory}. In these special cases the LV density variable $%
u^{(\alpha _{0})}(r,t)$ is proportional to the ordinary microscopic particle
density $u^{(\alpha _{0})}(r,t)=c^{(\alpha _{0})}\rho (r,t)$ with $\rho
(r,t)=\sum_{i=1}^{N}\delta (r-R_{i}(t))$ \cite{Mean}. Correspondingly for
these zero modes the components of the LV density correlation matrices $%
S^{(\alpha _{0}\alpha _{0})}(k)$ and $S^{(\alpha _{0}\alpha _{0})}(k,\omega )
$ are simply proportional to the static and dynamic structure factor $S(k)$
and $S(k,\omega )$, respectively. 
\begin{figure}[tbp]
\includegraphics*[scale=0.38]{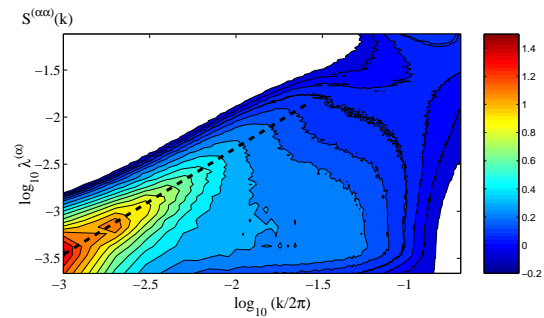}
\caption{Contour lines of the static correlation functions $S^{(\protect%
\alpha \protect\alpha )}(k)$ for a 1-d Lennard-Jones fluid ($N=100$, $L=1000$%
, $T=0.2$) plotted as function of $k$ and $\protect\lambda ^{(\protect\alpha %
)}$. The ridge at low $k-$ and $\protect\lambda -$values as indicated by the
dashed line, implies the existence of hydrodynamic Lyapunov modes.}
\label{fig:2}
\end{figure}

The general behavior of $S^{(\alpha \alpha )}(k)$ and $S^{(\alpha \alpha
)}(k,\omega )$ is inferred from numerical simulations. The particle dynamics
is obtained by integrating Hamilton's equations with periodic boundary
conditions using a standard truncated LJ interaction potential given by $%
V(r)=4\epsilon \lbrack (\sigma /r)^{12}-(\sigma /r)^{6}]-V_{c}$ for $%
r<r_{c}=2.5\sigma $ and $V(r)=0$ elsewhere \cite{Kob}. $V_{c}$ is chosen
such that $V(r)$ is continuous at $r_{c}$, and the parameters $\sigma $, $%
\epsilon $ and the particle mass $m$ are all set to unity. This means that
e.g. system length $L$, (kinetic) temperature $T$ and time is measured in
units of $\sigma $, $\epsilon $, and $(m\sigma ^{2}/48\epsilon )^{1/2}$,
respectively. The Lyapunov vectors $e^{(\alpha )}(t)$ were obtained
iteratively with the standard method \cite{EckmannRuelle}. The time step
used in the symplectic Verlet algorithm of our MD simulations is $\Delta
=0.008$ and the re-orthogonalization for the calculation of the $e^{(\alpha
)}(t)$ is repeated in time intervals of $30-100\Delta $. The numerical
evaluation of the correlation functions is started only after appropriate
equilibration periods. The coordinate part of a Lyapunov vector, as shown in
Fig.1 for $L=1000$ and $T=0.2$ is a quantity fluctuating in space and time.
The equal time correlations in space are captured by $S^{(\alpha \alpha )}(k)
$. For simplicity we present first results for the one-dimensional LJ fluid,
for which $S^{(\alpha \alpha )}(k)$ is a scalar quantity. In this case we
can depict its behavior in a contour plot. Since for $\lambda ^{(\alpha )}>0$
the latter decreases monotonically with the index $\alpha $, we can plot $%
S^{(\alpha \alpha )}(k)$ also over the $\lambda -k-$plane thereby
introducing the function $S(\lambda ,k)$. The surmise, which is supported by
our numerical simulations, is that the latter converges with increasing $N$
(for fixed temperature $T$ and density $N/L$) to a well-defined limit. In
the contour plot of Fig.2 we recognize a ridge at small $\lambda $ and $k$,
which defines a dispersion relation $\lambda (k)$ or $k(\lambda )$ implying
that the Lyapunov mode corresponding to Lyapunov exponent $\lambda $ is
dominated by a spatial oscillation with wave number $k(\lambda )$. In Fig.3
we determined the dispersion relation $\lambda (k)$ by plotting $\lambda $
versus $k_{\max }=\arg \max_{k}S(\lambda ,k)$ for various systems differing
in density and temperature. We find that the dispersion $\lambda (k)$
appears to be independent of these parameters in the regions where $\lambda
(k)$ is well defined. Only the extent of the latter regions is parameter
dependent, as it decreases e.g. with decreasing density. The common
dispersion relation $\lambda (k)$ being linear on a log-log scale results in
a power law $\lambda (k)\sim k^{\eta }$ with $\eta =1.2\pm 0.1$ although a
linear dispersion with quadratic corrections cannot be excluded. In any case
these results unambiguously show the existence of hydrodynamic Lyapunov
modes in LJ fluids. Actually, our investigations show that their existence
is not restricted to 1-d systems. In isotropic fluids with $d>1$ the
Cartesian components $F_{\mu \nu }^{(\alpha \alpha )}(k,\tau )$ of $%
F^{(\alpha \alpha )}(k,\tau )$ can be written in terms of longitudinal and
transverse correlation functions $F_{L}^{(\alpha \alpha )}$ and $%
F_{T}^{(\alpha \alpha )}$ as $F_{\mu \nu }^{(\alpha \alpha )}(k,\tau )=\hat{k%
}_{\mu }\hat{k}_{\nu }F_{L}^{(\alpha \alpha )}(\left| k\right| ,\tau
)+(\delta _{\mu \nu }-\hat{k}_{\mu }\hat{k}_{\nu })F_{T}^{(\alpha \alpha
)}(\left| k\right| ,\tau )$ with $\hat{k}_{\mu }=(k/\left| k\right| )_{\mu }$%
. We find for $d=2$ that $F_{L}^{(\alpha \alpha )}(\left| k\right| ,0)$ and $%
F_{T}^{(\alpha \alpha )}(\left| k\right| ,0)$ behave similar to the 1-d case
shown in Fig.2 implying the existence of hydrodynamic Lyapunov modes also
there. 
\begin{figure}[tbp]
\includegraphics*[scale=0.26]{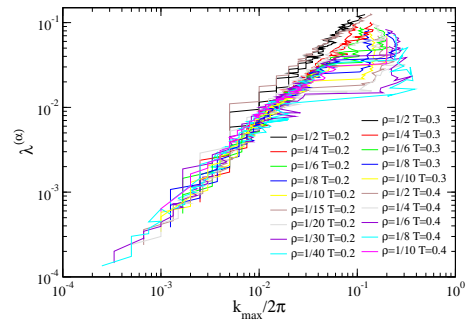}
\caption{The dispersion relation $\protect\lambda (k)$ of the hydrodynamic
Lyapunov modes shown for different particle densities and temperatures,
appears to be independent of the latter for a wide range of parameters.}
\label{fig:3}
\end{figure}

\begin{figure}[tbp]
\includegraphics*[scale=0.26]{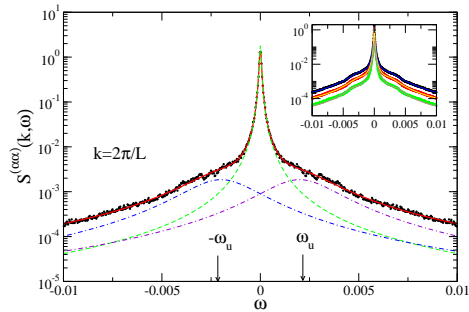}
\caption{The dynamic correlation function $S^{(\protect\alpha \protect\alpha %
)}(k,\protect\omega )$ for $\protect\alpha =96$ and $k=2\protect\pi /L$. The
full line results from a fit by a 3-pole approximation. The corresponding
decomposition into three Lorentzians is also shown. The insert shows a
comparison of such fits for $k=2\protect\pi /L$, $4\protect\pi /L$, and $8%
\protect\pi /L$ (from the top).}
\label{fig:4}
\end{figure}
More detailed information can be obtained from the dynamical LV correlation
functions $S^{(\alpha \alpha )}(k,\omega )$, which encode in addition to the
structural also the temporal correlations. As usual the equal time
correlations can be recovered by a frequency integration $S^{(\alpha \alpha
)}(k)=\int S^{(\alpha \alpha )}(k,\omega )~d\omega $. In Fig.4 we show
typical examples for $S^{(\alpha \alpha )}(k,\omega )$. It consist of a
central ''quasi-elastic'' peak with shoulders resulting from dynamical
excitations quite similar to the dynamical structure factor $S(k,\omega )$
of fluids. In order to extract the dynamical information we used a 3-pole
approximation for the Laplace transform of $F^{(\alpha \alpha )}(k,\tau )$
from Eq.(\ref{intermediate}), which amounts to fitting $S^{(\alpha \alpha
)}(k,\omega )$ by a superposition of three Lorentzians, one central peak at $%
\omega =0$ and two symmetric peaks located at $\omega =\pm \omega (k)$. The
fits are also shown in the figure and describe the frequency dependence of $%
S^{(\alpha \alpha )}(k,\omega )$ quite well. Such a dependence arises
naturally e.g. from continued fraction expansions based on Mori-Zwanzig
projection techniques \cite{GeneralizedHydrodynamics}, which as mentioned
may be applied also to this problem. These fits allow us to extract the
dispersion relations $\omega ^{(\alpha )}(k)$ for each of the hydrodynamic
Lyapunov modes with index $\alpha $. The result is shown in Fig.5 for
several of the Lyapunov modes. Clearly, this tells us that a Lyapunov mode
corresponding to exponent $\lambda $ is characterized, apart from the
dominating wave length $k(\lambda )$, by a typical frequency $\omega
(k(\lambda ))$. Because $\frac{d\omega }{dk}$ is non-vanishing, this implies
propagating wave-like excitations. The origin of the characteristic
frequency $\omega (k(\lambda ))$ is not yet fully understood. Probably it
reflects rotational motion of the orthogonal frame $\left\{ e^{(\alpha
)}(t)\right\} $ around its reference trajectory \cite{PoschForster}. The
full LV dynamics, however, is more complex. We find, for instance, that the
coherent wave-like motion is switched on and off intermittently \cite%
{YangRadons}. This phenomenon presumably is responsible for the strong
central peak in $S^{(\alpha \alpha )}(k,\omega )$ (see Fig.4).

In summary, we have shown that the LV density correlation functions provide
a valuable tool for quantifying spatiotemporal phenomena in many-particle
Lyapunov vectors. They allowed us to identify for the first time
hydrodynamic Lyapunov modes and their dynamics in soft-potential systems.
These findings indicate that the existence of hydrodynamic Lyapunov modes is
a universal phenomenon in chaotic translation invariant many-particle
systems, independent of the interaction potential. This view is supported by
the fact that these modes appear to be present also in one- and
two-dimensional fluids with Weeks-Chandler-Anderson (WCA) interaction
potentials \cite{PoschPriv}. There exist obvious generalizations of our
correlation functions, which will be treated in forthcoming papers. For
instance, one can consider correlations between different Lyapunov vectors
resulting in cross-spectra $S^{(\alpha \beta )}(k,\omega )$, or correlations
of the LV density not only in coordinate space but in phase space $%
e^{(\alpha )}(r,p,t)=\sum_{i=1}^{N}e_{i}^{(\alpha )}(t)~\delta
(r-R_{i}(t))\delta (p-P_{i}(t))$, which is an approach used similarly in
kinetic theory \cite{GeneralizedHydrodynamics}. 
\begin{figure}[tbp]
\includegraphics*[scale=0.26]{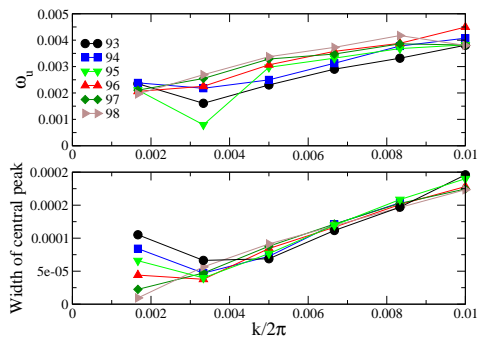}
\caption{Dispersion relations $\protect\omega ^{(\protect\alpha )}(k)$ (top)
and the $k-$dependence of the width of the central peak (bottom) for
different Lyapunov vectors obtained from 3-pole approximations as in Fig.4.
Note the clear separation of the corresponding time scales.}
\label{fig:5}
\end{figure}

\bigskip

\noindent \textbf{Acknowledgments} \bigskip \newline
We thank H. Posch, A. Pikovsky, W. Just, A. Latz, and especially W. Kob for
illuminating discussions. Support from the Deutsche Forschungsgemeinschaft
within the SFB 393 ''Numerische Simulation auf massiv parallelen Rechnern''
is gratefully acknowledged. 


\begin{thebibliography}{99}
\bibitem{Mean} In these cases the dynamical variables $u^{(\alpha )}$ have
non-vanishing mean values which have to be subtracted before the correlation
functions are formed.
\end{thebibliography}
\end{document}